\documentclass[twocolumn]{aastex631}

\usepackage{booktabs}
\usepackage{amsmath}

\begin{document}

\title{Tidal Disruption Event Demographics in Supermassive Black Hole Binaries Over Cosmic Times}

\author[0000-0002-7854-1953]{Denyz Melchor}
\affiliation{Department of Physics and Astronomy, University of California, Los Angeles, CA 90095, USA}
\affiliation{Mani L. Bhaumik Institute for Theoretical Physics, Department of Physics and Astronomy, UCLA, Los Angeles, CA 90095, USA}
\author[0000-0002-9802-9279]{Smadar Naoz}
\affiliation{Department of Physics and Astronomy, University of California, Los Angeles, CA 90095, USA}
\affiliation{Mani L. Bhaumik Institute for Theoretical Physics, Department of Physics and Astronomy, UCLA, Los Angeles, CA 90095, USA}

\author[0000-0003-3703-5154]{Suvi Gezari}
\affiliation{Department of Astronomy, University of Maryland, College Park, MD, 20742-2421, USA}
\affiliation{William H. Miller III Department of Physics and Astronomy, Johns Hopkins University, Baltimore, MD 21218, USA}

\author[0000-0001-6350-8168]{Brenna Mockler}
\affiliation{The Observatories of the Carnegie Institution for Science, Pasadena, CA 91101, USA}
\affiliation{Department of Physics and Astronomy, University of California, Davis, CA 95616, USA}

\begin{abstract}

Tidal disruption events (TDEs) offer a unique probe of supermassive black hole (SMBH) demographics, but their observed rates remain difficult to reconcile with standard single-SMBH models. In this work, we use simulations of SMBH binaries, including the combined effects of eccentric Kozai-Lidov oscillations and two-body relaxation, to explore how TDE rates scale with SMBH mass and redshift. We find that binary systems exhibit increasing TDE rates with mass, in contrast to the declining trend expected for single SMBHs. These binary-driven rates match those observed in post-starburst galaxies, suggesting that a subset of TDE hosts may contain SMBH binaries. TDE light curves in some massive galaxies exhibit unexpectedly short durations, suggesting that the disrupting SMBH may be less massive than implied by host galaxy scaling relations, consistent with disruptions by the less massive black hole in a binary. By convolving our mass-dependent rates with the SMBH mass function, we predict redshift-dependent TDE rates, which we show can be used to constrain the supermassive black hole binary fraction. Our results provide a testable framework for interpreting TDE demographics in upcoming wide-field surveys such as LSST and Roman.
\end{abstract}


\section{Introduction} \label{sec:intro}

\begin{figure*}
   \begin{center} 
    \includegraphics[width=0.8\textwidth]{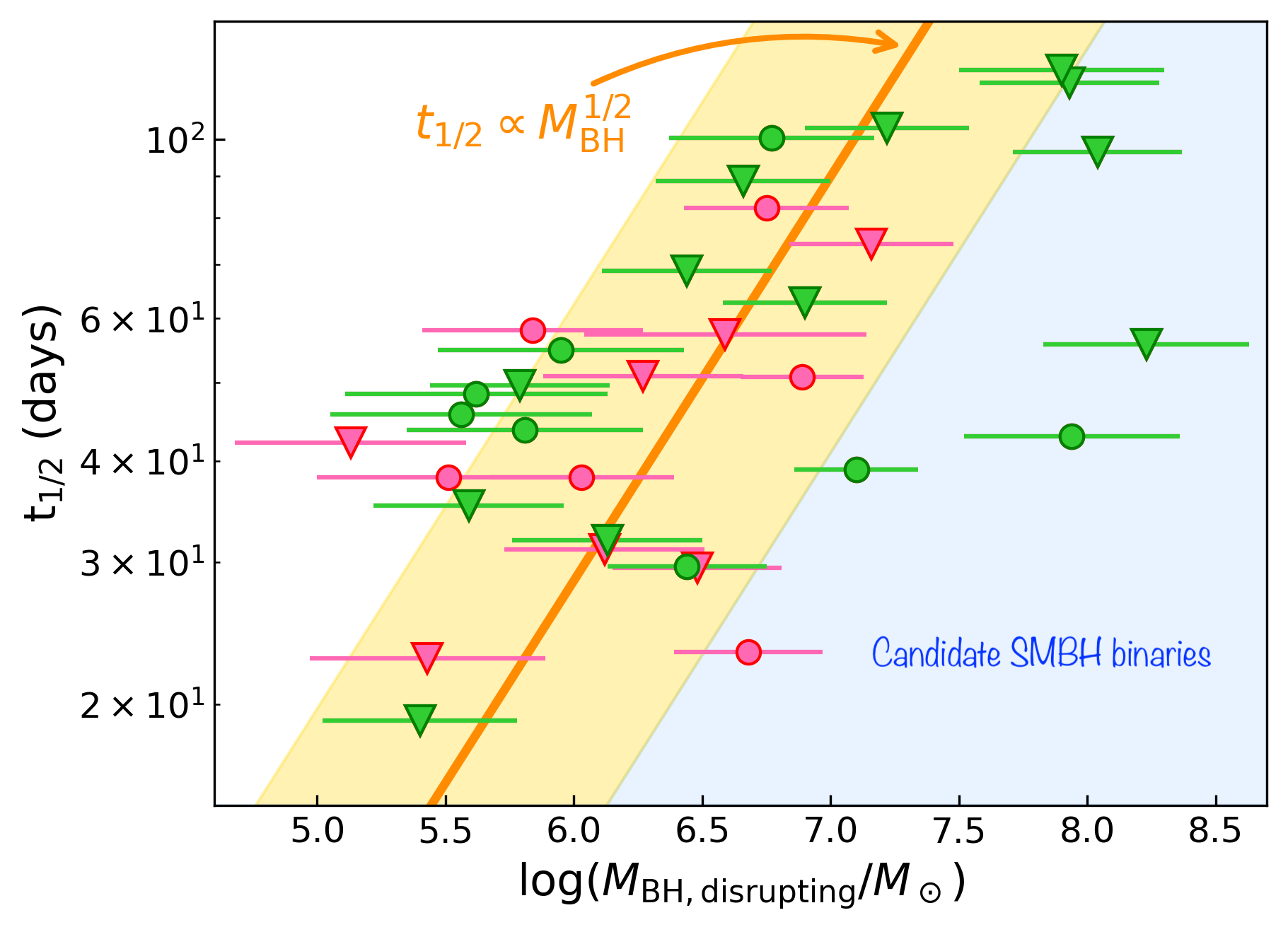} 
  \end{center} 
  \caption{  \upshape {\bf Correlation between TDE light curves and black hole mass. } Adapted from \cite{yao_tidal_2023}, this figure depicts the relationship between black hole masses (estimated from host galaxy properties; x-axis) and t$_{1/2}$, the duration a TDE remains above half its peak luminosity (rise plus decline; y-axis). Green markers denote TDEs in green valley galaxies, while the pink markers fall outside that category.  The orange curve is a best-fit relation with fixed slope $t_{1/2} \propto \rm{M}_{BH}^{1/2}$, and the yellow shaded band shows a 1$\sigma$ scatter.  Circle markers use black hole masses estimates from M$_{BH}-$M$_{\rm gal}$; inverted triangles use estimates from the M$_{BH}- \sigma_\star$ relation. The shaded area below the lower 1$\sigma$ bound highlights candidate TDEs that may originate from SMBH binary systems. }
  \label{fig:Mh_vs_Mh}   
\end{figure*}


Supermassive black holes (SMBHs) are central to galaxy evolution, regulating star formation and shaping the large-scale structure of the Universe \citep[e.g.,][]{silk_quasars_1998, di_matteo_energy_2005, springel_modelling_2005, bower_breaking_2006, hopkins_cosmological_2008, fabian_observational_2012, kormendy_coevolution_2013, heckman_coevolution_2014}. Most massive galaxies, including the Milky Way, host a central SMBH embedded in a dense stellar cluster \citep[e.g.,][]{ghez_high_1998, ghez_galactic_2009}. Galaxy mergers naturally lead to SMBH binaries \citep[e.g.,][]{begelman_massive_1980, di_matteo_energy_2005, hopkins_unified_2006, robertson_mergerdriven_2006, callegari_pairing_2009, li_pairing_2020}, which can remain bound for $\sim$~Gyr timescales, dynamically influencing their environments before coalescence \citep[e.g.,][]{kelley_massive_2017}.

However, SMBH binaries remain difficult to observe across most evolutionary stages. While wide dual AGN systems have been identified at kiloparsec scales \citep[e.g.,][]{komossa_observational_2003, bianchi_chandra_2008, comerford_175_2009, comerford_origin_2018, green_sdss_2010, smith_search_2010, liu_discovery_2010, foord_second_2020, li_pairing_2020, stemo_catalog_2021}, and a few parsec-scale systems show variability or radio signatures \citep[e.g.,][]{sillanpaa_oj_1988, rodriguez_compact_2006, komossa_recoiling_2008, bogdanovic_tidal_2004, dotti_supermassive_2007}, intermediate-separation binaries remain challenging to detect. Current observational techniques such as AGN variability monitoring, radio interferometry, and pulsar timing arrays \citep[e.g.,][]{afzal_nanograv_2023, agazie_nanograv_2023} offer only partial constraints, due to limited sample sizes, sensitivity, and contamination by false positives.
Future observatories like the Laser Interferometer Space Antenna (LISA) are expected to offer key insights into the final stages of binary evolution by detecting gravitational waves from merging binaries in the low-frequency regime \citep[e.g.,][]{amaro-seoane_laser_2017, baker_laser_2019}. Constraining the binary fraction and its redshift evolution is vital for understanding SMBH growth, galaxy assembly, and the gravitational wave background \citep[e.g.,][]{volonteri_assembly_2003, sesana_interaction_2006, menou_merger_2001, haiman_population_2009, amaro-seoane_laser_2017, baker_laser_2019}.

Given these challenges, we propose to infer SMBH binarity through the populations of tidal disruption events (TDEs). A TDE occurs when a star is torn apart by an SMBH’s tidal forces upon approaching within a characteristic tidal radius \citep[e.g.,][]{hills_possible_1975, rees_tidal_1988, guillochon_hydrodynamical_2013}, 
\begin{equation}\label{eq:tidal_radius}
R_{\rm T} \approx r_\star \left(\frac{M_{\rm BH}}{m_\star}\right)^{1/3} \ ,
\end{equation}
where $r_\star$ and $m_\star$ are the stellar radius and mass, and $M_{\rm BH}$ is the SMBH mass. The resulting accretion flare can temporarily illuminate otherwise quiescent SMBHs \citep[][]{evans_tidal_1989, ulmer_flares_1999, guillochon_ps1-10jh_2014}.


The presence of a secondary SMBH has been shown to enhance TDE rates. Perturbations from the companion can increase the rate of stellar disruptions by increasing loss-cone refilling through the eccentric Kozai-Lidov mechanism \citep[EKL; e.g.,][]{kozai_secular_1962, lidov_evolution_1962, naoz_eccentric_2016}, while gravitational interactions within the surrounding stellar population also contribute \citep{chen_enhanced_2009, naoz_combined_2022, melchor_tidal_2024}. Previous studies have proposed using TDE rates, and even future detections of extreme mass ratio inspiral (EMRI) events, to constrain the SMBH binary fraction \citep{naoz_combined_2022, naoz_enhanced_2023, melchor_tidal_2024}. 

This combined binary mechanism may also help explain the excess of TDEs observed in post-starburst (PSB) galaxies, which show TDE rates up to two orders of magnitude higher than in the general galaxy population \citep[e.g.,][]{van_velzen_optical-ultraviolet_2020}. Several other mechanisms have been proposed to account for this enhancement, including changes to the stellar dynamical environment or stellar population. These include the presence of massive perturbers that can enhance loss cone refilling \citep[e.g.,][]{perets_massive_2007, perets_massive_2008}, interactions with AGN disks that funnel stars inward \citep[e.g.,][]{kennedy_stardisc_2016}, or evolutionary effects of a fading AGN disk  \citep[e.g.,][]{wang_explanation_2024}. Dynamical effects from eccentric nuclear disks have also been shown to boost disruption rates through coherent torques and increased orbital eccentricities \citep[e.g.,][]{madigan_dynamical_2018}. However, \citet{teboul_strong_2025} find that models invoking ultra-steep stellar density profiles, velocity anisotropies, or top-heavy present-day mass functions do not reproduce the observed strength or duration of the TDE excess in PSB galaxies. These limitations point to the need for additional mechanisms, such as the SMBH binary scenario explored in this work.


Most PSB galaxies fall in the green valley, a transitional region between the blue and red sequences in color-magnitude phase space, which may indicate a shift in star formation activity \citep[e.g.,][]{wong_galaxy_2012, alatalo_welcome_2017}. This shift is often associated with recent mergers, making PSBs promising environments for hosting SMBH binaries. If post-merger binaries are responsible for boosting TDE rates, these events offer a novel probe of SMBH binary demographics across time. While \cite{stone_delay_2018} argue that the elevated TDE rates in PSB galaxies cannot be explained by dynamical evolution alone, our combined framework involves the effects that binaries have on TDE rates, which were not incorporated in their analysis.


In this work, we quantify how TDE rates depend on SMBH mass and redshift in galaxies with SMBH binaries and compare our results with the single SMBH scenario as well as current TDE observations. We also present predictions for upcoming time-domain surveys, including the Vera C. Rubin Observatory's Legacy Survey of Space and Time (LSST) \citep[][]{ivezic_lsst_2019} and NASA’s Roman Space Telescope \citep[][]{spergel_wide-field_2015}.

This paper is organized as follows. In Section \ref{sec:smbh_bins}, we describe potential signatures of SMBH binaries in TDE light curves. Then, in Section \ref{sec:tde_rates}, we detail the combined effects of two-body relaxation and the eccentric Kozai-Lidov mechanism in producing TDEs in SMBH binaries, including how the TDE rate varies with disrupting SMBH mass and evolves with redshift. In Section \ref{sec:discussion}, we discuss observational implications of our findings. Additional modeling details are provided in the Appendix (Section \ref{sec:appendix}).

\section{SMBH Binary Signatures}\label{sec:smbh_bins}
\begin{figure*}
  \begin{center} 
    \includegraphics[width=0.9\textwidth]{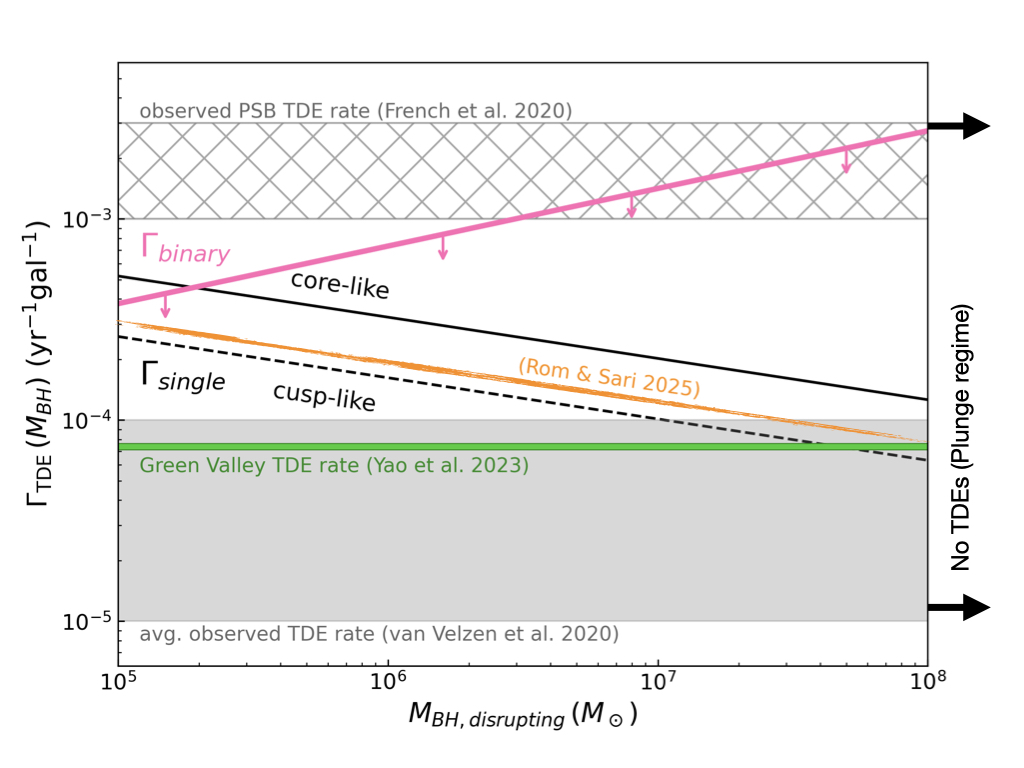}
  \end{center} 
  \caption{  \upshape {\bf Projected TDE rate per SMBH mass.} The pink curve shows the predicted upper-limit TDE rate as a function of disrupting SMBH mass for a binary with fixed mass ratio ($q=100$) and a core-like stellar density profile. The solid and dashed black lines show single-SMBH TDE rates from two-body relaxation assuming core-like and cusp-like profiles, respectively. The orange line represents the TDE rate estimated by \cite{rom_segregation_2025}. The hatched gray band shows the observed TDE rate in PSB galaxies \citep[][]{french_structure_2020}, while the solid gray band denotes the average observed TDE rate \citep[][]{van_velzen_optical-ultraviolet_2020}. The green line marks the TDE rate in green valley galaxies estimated by \cite{yao_tidal_2023}. Disruptions of Sun-like stars are not expected beyond M$_{\rm BH} \sim 10^8 \rm{M}_\odot$ \citep[e.g.,][]{stone_delay_2018, wevers_black_2019, gezari_tidal_2021, polkas_demographics_2024}, so a decline in the TDE rate is expected at higher masses.} \label{fig:rate_per_mass} 
\end{figure*}


TDEs are powerful tools for estimating SMBH masses. Hydrodynamical simulations combined with disk reprocessing models have shown that early-time light curve properties, like rise time and peak luminosity, correlate with the mass of the disrupting SMBH \citep{mockler_weighing_2019}. The late-time plateau luminosities (at t$_{\rm peak} \gtrsim 100$ days) have also been proposed as independent diagnostics of black hole mass \cite[][]{mummery_fundamental_2024}. Additionally, TDE light curve demographics reveal a flat SMBH mass function with a drop for $M_{{\rm BH}} \gtrsim 10^{7.5}$ M$~\odot$, due to direct stellar capture by more massive SMBHs \cite[][]{yao_tidal_2023}. These trends suggest that light curve properties can serve as indirect probes of SMBH mass.

The fallback timescale, $t_{\rm fb}$, is suggested to be one of the key observables in TDE light curves that encodes this mass dependence \citep[e.g.,][]{holoien_asassn-14ae_2014, auchettl_new_2017, blagorodnova_iptf16fnl_2017, hung_revisiting_2017, holoien_ps18kh_2019, mockler_weighing_2019, van_velzen_first_2019, gomez_tidal_2020}. In particular, this is the timescale for the most bound debris of a TDE to return to the black hole. For a star of mass $m_\star$ and radius $r_\star$ disrupted by an SMBH of mass $M_{\rm BH}$ at the tidal radius $R_T$, the fallback timescale is 

\begin{equation}\label{eq:TDEFB}
\begin{split}
t_{\mathrm{fb}} &\approx 2 \pi \frac{R_T^3}{r_\star^{3/2}} \frac{1}{\sqrt{G M_{\mathrm{BH}}}} 
= 4 \pi \frac{r_\star^{3/2}}{m_\star} \sqrt{\frac{M_{\mathrm{BH}}}{G}} \\
&\approx \left( \frac{M_{\mathrm{BH}}}{10^6 M_\odot} \right)^{1/2} m_\star^{-1} r_\star^{3/2} \; 41\, \mathrm{days} \, ,
\end{split}
\end{equation}
where the normalization (41 days) reflects a solar-type star disrupted by a black hole of $10^6$M$_{\odot}$, following \citet{yao_tidal_2023}.
Variations in the mass and radius of disrupted stars can introduce scatter in the fallback timescale, potentially obscuring its expected scaling with SMBH mass. 

An alternative timescale introduced in \cite{yao_tidal_2023} is $t_{1/2}$, defined as the duration over which the TDE light curve remains above half of its peak luminosity. Photometric data directly constrain this quantity and provide a useful measure of the flare's duration near maximum light. Although $t_{1/2}$ is not equivalent to $t_{fb}$, it is expected to retain some imprint of the system's dynamical properties. 

We adapt Figure 12e from \cite{yao_tidal_2023}, and show $t_{1/2}$ as a function of galaxy-based SMBH mass estimates in Figure \ref{fig:Mh_vs_Mh}. Green markers indicate TDEs hosted by galaxies falling within the green-valley constraints (as defined by Eq. 22 in \citealt{yao_tidal_2023}), while the pink markers correspond to TDEs outside this region in color-mass space. A best-fit relation to the data (orange line) is also shown, with the slope fixed to $t_{1/2} \propto M_{\rm BH}^{1/2}$, and the yellow shaded region denoting 1$\sigma$ scatter. 

The blue shaded region, below the best-fit curve, highlights systems where the TDE duration is shorter than expected based on galaxy-derived BH mass estimates. As seen, there is a significant scatter to the right of the fit, with comparatively narrower scatter to the left. This asymmetry further suggests that the galaxy-based estimates may systematically overestimate the mass of the disrupting SMBH in a subset of TDEs. 


This trend also appears in Figure \ref{fig:combined_tau_Mh} (see Appendix: Section \ref{sec:appendix}) which shows the exponential decay timescale, $\tau$, for TDEs from \cite{hammerstein_final_2022} (left panel) and \cite{van_velzen_optical-ultraviolet_2020} (right panel), plotted against black hole masses estimated from M$_{BH}-\sigma_\star$ or M$_{BH}-$M$_{\rm gal}$ found in \cite{mummery_fundamental_2024} for the same events (see Table \ref{tab:combined_tde_data}). A best-fit line (orange) with fixed slope, $\tau \propto M_{\rm BH}^{1/2}$, is shown here as well. Notably, even using these alternative light-curve measurements, we observe a consistent asymmetric distribution with wider scatter to the right compared to the left of the best-fit curve. 

This persistent asymmetry across different datasets and observables raises a key question: Is this pattern intrinsic to the TDE itself, or does it reveal a missing mechanism or observational bias that could explain the deviation from symmetry around the expected fallback timescale? 

A possible explanation is that these TDEs originate from the lower-mass component of an unresolved SMBH binary. In such systems, galaxy-based scaling relations may reflect the more massive SMBH of the binary, leading to systematic overestimates of the mass of the disrupting black hole. 
If some TDEs instead occur in binary systems, then the inferred SMBH mass would probe only one member of the binary. This could lead to wide discrepancies between TDE-derived masses and those estimated from host galaxy scaling relations (e.g., $M_{\rm BH} -\sigma_\star$), where in a binary, the more massive BH dominates the gravitational potential.  Recent work by \cite{mockler_uncovering_2023} supports this scenario, showing that TDEs around less-massive SMBHs can yield systematically underestimated masses compared to expectations from host galaxy properties, hinting at unresolved SMBH binaries.

If a subset of TDE-derived SMBH masses is systematically underestimated, it could imply a population of unresolved SMBH binaries in TDE host galaxies. To address this, our study quantifies how binary dynamics alter TDE rates across time, leveraging observable signatures to distinguish binary-driven disruptions from single SMBH events. These predictions will guide targeted searches with next-generation surveys \citep[e.g.,][]{spergel_wide-field_2015, ivezic_lsst_2019}, providing a constraint on the binary population through TDE demographics.



\section{TDE\lowercase{s} in SMBH binaries}\label{sec:tde_rates}
One of the promising mechanisms for TDE formation involves loss cone dynamics, which describes the scattering interactions between stars in a nuclear star cluster. This two-body relaxation arises from cumulative gravitational interactions between stars, gradually altering their orbits and driving them into the loss cone of the central black hole, resulting in a TDE  \citep[][]{frank_effects_1976, rees_tidal_1988, rauch_resonant_1996, linial_stellar_2022, rose_stellar_2023, rom_dynamics_2024}. 

In SMBH binary systems, we have previously demonstrated that the combined effects of EKL and two-body relaxation efficiently drive stars onto the SMBH, yielding TDEs and repeated TDEs \citep{melchor_tidal_2024}. In this setup, the companion SMBH exerts gravitational perturbations on the stellar cluster surrounding the disrupting black hole, altering stellar trajectories. As in \cite{melchor_tidal_2024}, we set the primary SMBH to be less massive than the companion, a configuration shown to enhance TDE rates by avoiding general relativistic precession that would otherwise suppress the high-eccentricity excitations necessary for stellar disruptions \citep[see also,][]{mockler_uncovering_2023}. 

Estimating the TDE rate involves making assumptions about the system's density profile distribution, typically characterized as either core-like (e.g., the index power-law, $\alpha=1$) or cusp-like ($\alpha=2$), along with the number of particles and other parameters. A steady-state assumption is often used, although not always justified. Nevertheless, this assumption is employed for both single and binary SMBH cases. We calculate the single SMBH rate within the SMBH sphere of influence, $R_{sph}$, following standard formalism (see Equation \ref{eq:gamma_rlx}, in Appendix \ref{sec:appendix}). For SMBH binaries, we estimate the upper limit TDE rate by following the framework from \cite{naoz_combined_2022} and \cite{melchor_tidal_2024}, modeling systems under the combined influence of two-body relaxation and the EKL mechanism.

To determine the impact of SMBH binaries on TDE rates, we first calculate the TDE rate as a function of the disrupting SMBH mass, comparing binary systems to single SMBH scenarios. Theoretical models predict a linear scaling between TDE rate and the disrupting SMBH mass \cite[][]{naoz_enhanced_2023, mockler_uncovering_2023, rom_segregation_2025}. Figure \ref{fig:rate_per_mass} compares the simulated upper limit of the TDE rate as a function of the disrupting mass for SMBH binaries (pink line) to theoretical predictions for single SMBHs with core-like (solid black) and cusp-like (dashed black) stellar distributions. We also depict the TDE rate from strong mass segregation, adopted from \cite{rom_segregation_2025} (orange), who model a broken power-law for the present-day stellar mass function to calculate two-body relaxation-driven rates.


To contextualize these rates with observations, Figure \ref{fig:rate_per_mass} also includes the average observed TDE rate \citep[gray band,][]{van_velzen_optical-ultraviolet_2020}, the TDE rate in PSB galaxies \citep[hatched gray band,][]{french_structure_2020}, and more broadly, the observed green valley TDE rate \citep[green line,][]{yao_tidal_2023}. TDEs are not expected for Sun-like stars around SMBHs with masses near $10^8 \rm{M}_\odot$, as the tidal radius falls within the event horizon, resulting in direct stellar plunges \citep[e.g.,][]{kesden_tidal_2012}. Our models do not account for this plunge effect in either the single or binary SMBH scenarios. Nonetheless, we expect the TDE rate to decline at $\sim 10^8 \rm{M}_\odot$ consistent with both theoretical predictions and observational evidence \citep[e.g.,][]{stone_delay_2018, wevers_black_2019, gezari_tidal_2021, polkas_demographics_2024}.

Our results suggest that binaries produce TDEs at a higher rate across a broad range of SMBH masses, particularly for more massive SMBHs, where our rate approaches that observed in PSB galaxies. The elevated TDE rates observed in PSBs may, therefore, serve as indirect evidence of unresolved SMBH pairs.

As mentioned, post-starburst galaxies are shown to be overrepresented among TDE host galaxies \citep[e.g.][]{french_tidal_2016, hammerstein_tidal_2021}. Given that PSB galaxies often result from recent mergers, it is reasonable to suggest that they may host SMBH binaries formed during galactic interaction \citep[][]{haehnelt_correlation_2000}. The agreement between our simulated binary TDE rates and the observed PSB rates suggests that SMBH binaries could be a common feature in PSB galaxies and may explain why TDEs are more frequently detected in such environments compared to the general galaxy population.



The connection between SMBH binaries, PSB host galaxies, and elevated TDE rates may suggest that the evolving TDE rate across redshift could mirror the merger history of SMBHs. 
To model the redshift evolution of TDEs, we specifically convolve the TDE rate per SMBH mass, $\gamma(M)$, with the SMBH mass function from \cite{merloni_synthesis_2008}, yielding the volumetric rate, 
 \begin{equation}
    \Gamma (z) = \int\frac{d^2 N}{dV d\ln M} \gamma (M) d\ln M \ , 
\end{equation}
where ${d^2 N}/{dV d\ln M}$ is the SMBH number density. We limit our analysis to $z \leq 2$, corresponding to the epoch of peak star formation and galaxy merger activity, when SMBH growth and binary activity are most prominent \citep[e.g.,][]{madau_cosmic_2014}.
 The resulting number of TDEs per unit redshift, accounting for cosmological time dilation, $(1+z)^{-1}$, and co-moving volume, ${dV}/{dz}$, is then 
\begin{equation}
    \frac{dN}{dz}(z) =  \frac{\Gamma (z)}{1+z}\frac{dV}{dz} \ .
\end{equation}

\begin{figure*}
   \begin{center} 
    \includegraphics[width=\textwidth]{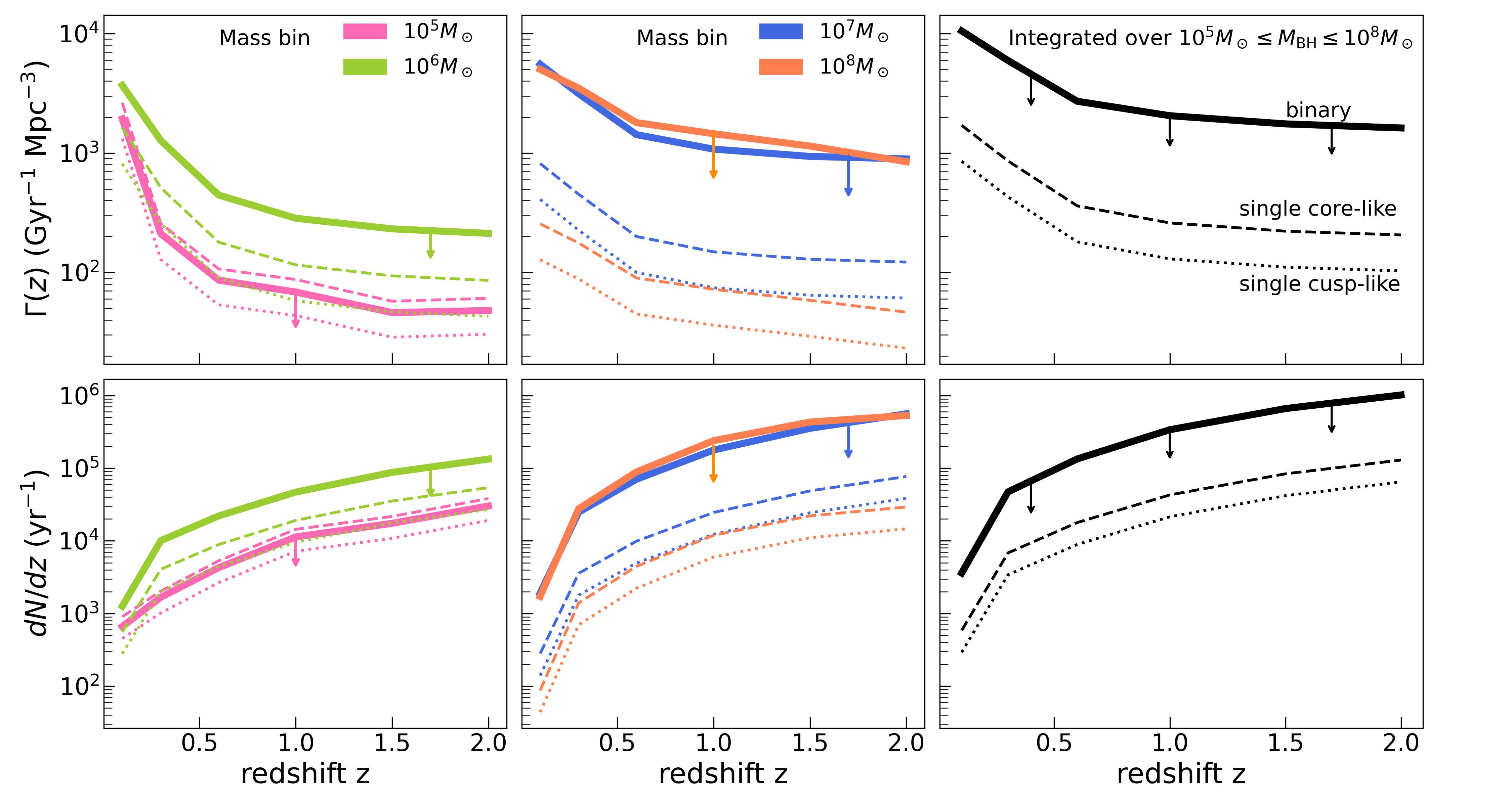} 
  \end{center} 
  \caption{  \upshape {\bf Upper limit TDE redshift evolution. } \textbf{Top panel:} The comoving volumetric TDE rate as a function of redshift $z$. The predicted rate from SMBH binaries (thick solid line) is compared to single-SMBH rates assuming core-like (dashed line) and cusp-like (dotted line) stellar profiles. The rightmost panel shows the rate integrated over SMBH masses from $10^5$ to $10^8, \ M_\odot$. The left and center panels show rates in individual mass bins: $ 10^5 M_\odot$ (pink) and $10^6 M_\odot$ (green) in the left panel, and $10^7 M_\odot$ (blue) and $10^8 M_\odot$ (orange) in the center panel. \textbf{Bottom panel:} The cumulative number of TDEs per unit redshift is shown comparing the binary (solid) and single SMBH (core-like: dashed; cusp-like: dotted) models. Mass bins follow the same color scheme as in the top panel. }
  \label{fig:rate_per_redshift}   
\end{figure*}


Motivated by this potential link, the top panels of Figure \ref{fig:rate_per_redshift} show this volumetric TDE rate, $\Gamma (z)$, evolving with redshift. The SMBH binary channel upper limit estimation (solid line) predicts rates several orders of magnitude higher than those of two-body relaxation alone (core-like, dashed; cusp-like, dotted). The bottom panel of the figure shows the cumulative number of TDEs as a function of redshift, ${dN}/{dz}$. 

In the central and rightmost panels of Figure \ref{fig:rate_per_redshift}, we observe that the single-SMBH rate at $10^7\, \rm{M}_\odot$ exceeds that of $10^8\, \rm{M}_\odot$, consistent with the expected decline in disruption efficiency at higher masses due to direct stellar plunges. In contrast, the binary-driven rates show only a modest difference between $10^7\, \rm{M}_\odot$ and $10^7\,\rm{M}_\odot$, reflecting the EKL mechanism's ability to continue fueling disruptions even in regimes where two-body relaxation becomes inefficient.

These findings reveal two key trends. First, the TDE rate as a function of disrupting SMBH mass for SMBH binaries increases with the mass of the disrupting black hole, contrasting the decreasing trend observed for single-SMBH systems. This divergence offers a potential observational signature for identifying binary SMBHs based on the mass-dependent TDE rate. Second, the combined EKL and two-body relaxation mechanism produces significantly more TDEs per unit redshift compared to two-body relaxation alone. This enhancement suggests that SMBH binaries could contribute more substantially to the observed TDE population than previously recognized. In all, these results are particularly relevant for PSB galaxies, where elevated TDE rates may signal the presence of unresolved SMBH binaries.

\section{Discussion}\label{sec:discussion}
Understanding the connection between TDE light curves and SMBH properties offers a powerful probe of black hole demographics and dynamics. We find a subset of TDEs with light curve timescales that are unexpectedly short given the SMBH masses inferred from host galaxy scaling relations.  
Specifically, as shown in Figures \ref{fig:Mh_vs_Mh} and \ref{fig:combined_tau_Mh},  the SMBH masses implied by a log-linear fit to $t_{1/2}$, which scales approximately as $M_{\rm BH}^{1/2}$, are systematically lower than those estimated from the host galaxies. 
One possible explanation is that some of these TDEs, which appear to be associated with high-mass SMBHs, are instead disrupted by the less massive member of an SMBH binary. 
 In this scenario, the host’s scaling relations reflect the more massive, non-disrupting black hole, leading to an overestimate of the actual disruptor mass. 
The consistency of this trend across both $t_{1/2}$ and the exponential decay time (${\tau}$) is suggestive of a binary SMBH origin in some TDEs.

Our previous work, \cite{melchor_tidal_2024}, demonstrated that the combined effects of EKL and two-body relaxation can significantly enhance TDE rates in SMBH binaries relative to single SMBH systems. We extended our analysis to explore a range of disrupting SMBH masses. By varying the mass of the disrupting black hole, our simulations of TDEs in SMBH binaries reveal a dichotomy in how TDE rates scale with SMBH mass. In binary systems, the TDE rate increases with SMBH mass due to EKL-driven stellar disruptions, while in single SMBHs, the rate declines with mass. We note that while our TDE rate increases with SMBH mass in our binary model, both the binary and single SMBH rates are expected to decline beyond M$_{\rm BH} \sim 10^8 \, \rm{ M}_\odot$, consistent with trends seen in other theoretical studies. For instance, \citet{polkas_demographics_2024} reports a sharp decrease in rates near this mass, in line with observational findings that suggest an intrinsic cutoff at high SMBH masses \citep[e.g.,][]{wevers_black_2019, gezari_tidal_2021}.

Notably, the TDE rates from the binary model match those observed in PSB galaxies for $M_{\rm BH} \gtrsim$ few $\times \ 10^6 M_\odot$ (Figure \ref{fig:rate_per_mass})\footnote{Note that recent studies propose that TDE rates in single SMBHs follow a broken or narrow power-law distribution as a function of mass, rather than a monotonic decline, due to mechanisms such as stellar anisotropy or mass segregation \citep[e.g.,][]{stone_rates_2016, rom_dynamics_2024, chang_rates_2024, rom_segregation_2025}. A more detailed comparison incorporating these models is left for future work. }. Although PSBs are a subset of green valley galaxies, their likely merger-driven histories create distinct conditions favorable to TDEs, such as high central stellar densities \citep[e.g.,][]{french_host_2020, french_tidal_2016}. The agreement between our binary model rate and the observed PSB rate, 
suggests that PSB TDE hosts may contain SMBH binaries, specifically hosting a disrupting SMBH with mass ranging between a $\rm{few \,} \times 10^6 - 10^8 \rm{M}_\odot $.


Building on this, we examine 
how our binary model scales across redshift, and convolve our mass-dependent TDE rates from Figure \ref{fig:rate_per_mass} with the SMBH mass function from \cite{merloni_synthesis_2008}. 
The resulting redshift-dependent, volumetric TDE rates, shown as upper limits in Figure \ref{fig:rate_per_redshift}, highlight two key trends. In the upper panels, the volumetric rate decreases with redshift, reflecting the declining number density of SMBHs at earlier cosmic times. In contrast, the lower panels show that the total number of TDEs, obtained by integrating over SMBH mass, increases with redshift due to the larger comoving volumes being sampled. Together, these curves represent theoretical upper bounds on TDE rates from SMBH binaries across redshift.

These upper limits complement and contextualize previous TDE rate estimates. For example, 
\citet{stone_rates_2016}, do not include the contribution of nuclear star clusters (NSCs), and therefore represent lower limits. In contrast, the semi-analytic model from \cite{polkas_demographics_2024} demonstrates that NSCs substantially contribute to the overall TDE rate. Their results indicate that systems with higher central densities dominate the disruption budget. Observational work by \cite{french_tidal_2016} further suggests that the elevated TDE rates in PSB galaxies are more likely due to compactness than to variations in the initial stellar mass function \citep[e.g.,][]{bortolas_tidal_2022}. Taken together, these theoretical and observational results underscore the importance of compact stellar environments in shaping TDE demographics.
This supports the plausibility of our binary-induced upper limits: SMBH binaries, especially in the aftermath of gas-rich mergers or starbursts, are expected to reside in dense, compact environments that naturally enhance the TDE rate.

The TDE rate for single SMBHs generally declines at higher black hole masses due to an increase in direct stellar plunges, which bypass tidal disruption altogether \citep[e.g.,][]{brockamp_tidal_2011, kesden_tidal_2012, merritt_loss-cone_2013}.
As noted earlier, incorporating plunges into our framework would introduce a similar suppression in the single and binary-driven rate around $10^8~$M$_\odot$. However, our central result remains robust: the TDE rate increases with the mass of the disrupting SMBH in binary systems. This trend arises because the combined influence of the EKL mechanism and two-body relaxation continues to efficiently drive stars into the loss cone in regions of parameter space where two-body relaxation alone is insufficient. This extended mass reach is a key distinguishing feature of the binary disruption channel, enabling TDEs to occur in mass regimes where the single-SMBH rate sharply declines.

To incorporate uncertainties in our model and the diversity of SMBH binary properties, we introduce a tuning parameter, $\eta$, where $\eta = 1$ corresponds to the assumption that all TDEs arise from our combined binary mechanism (EKL + two-body relaxation), and $\eta \rightarrow 0$ represents the limiting case in which binaries contribute negligibly. This parameter encapsulates uncertainties in the fraction of systems hosting SMBH binaries, binary separations and eccentricities, stellar densities within the sphere of influence, gas content, and other poorly constrained astrophysical factors. Our simulations assume maximal contribution from binary-driven processes, where each SMBH retains its surrounding stellar cluster post-merger and maintains a steady-state stellar population with ongoing replenishment.

\begin{figure*}
  \begin{center} 
    \includegraphics[]{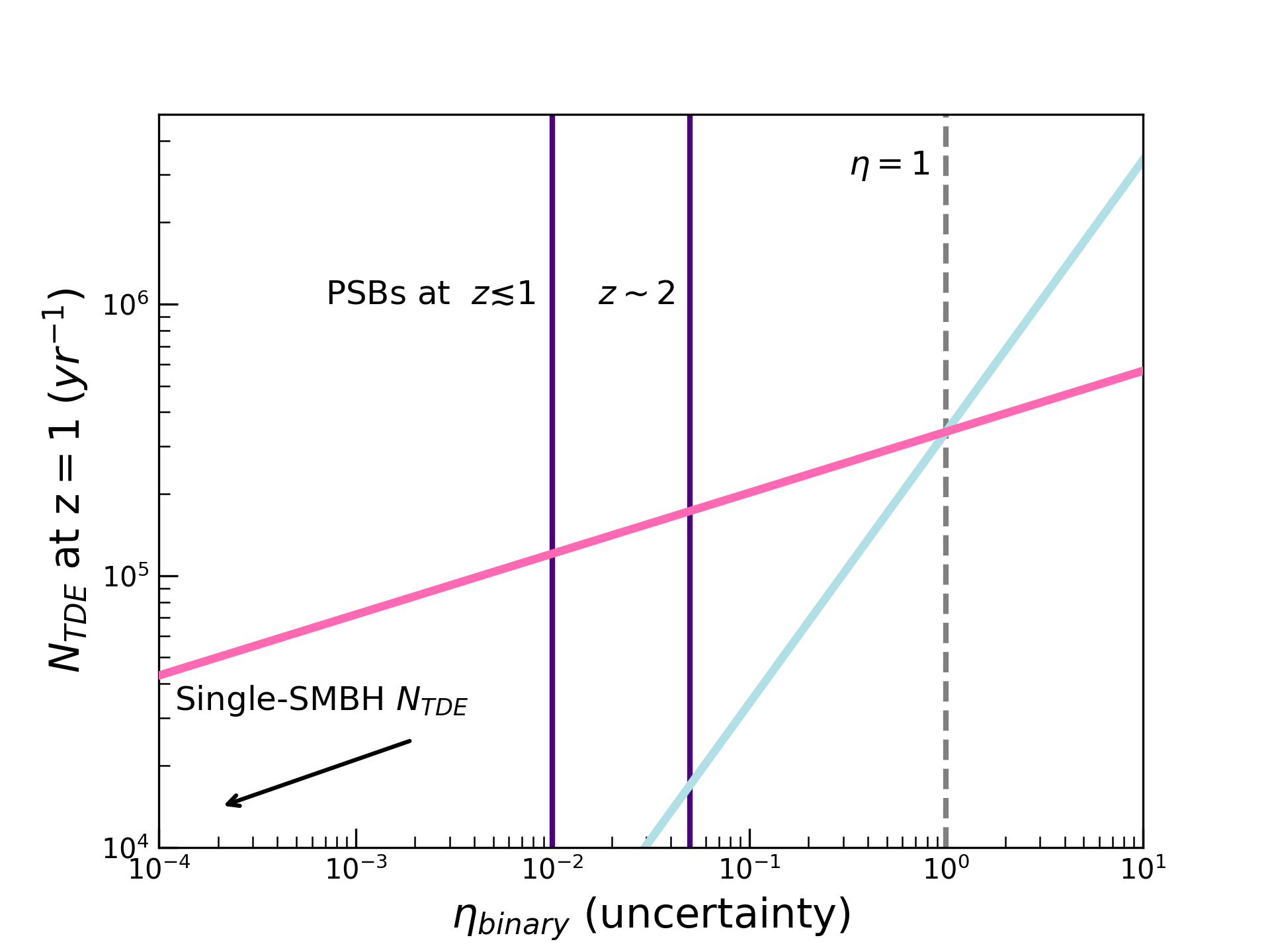}
  \end{center} 
  \caption{ \upshape {\bf Number of TDEs from an SMBH binary: Uncertainty at $z = 1$.} The x-axis shows $\eta$, which represents the confidence in the binary mechanism (EKL + two-body relaxation) producing TDEs, and the y-axis gives the predicted TDE count per redshift at $z=1$. The vertical dashed gray line at $\eta = 1$ indicates full confidence in binary-driven origins of TDEs. The intersection of the pink curve and the light blue line represents the predicted TDE count under this assumption. As $\eta$ $\rightarrow 0$, the pink curve approaches the single-SMBH TDE rate. The light blue provides a linear extrapolation of uncertainty anchored at $\eta = 1$. Vertical purple lines mark the observed fraction of PSB galaxies at $z \lesssim 1$ and $ z \sim 2$; their intersections with the model lines correspond to the expected TDE contribution from binaries in PSB hosts at those redshifts.}\label{fig:uncertainty} 
\end{figure*}

This is illustrated in Figure \ref{fig:uncertainty}, which shows how TDE rates vary with $\eta$. The pink curve shows our predicted TDE count at $z=1$, which approaches the single-SMBH baseline 
as $\eta \rightarrow 0$. To facilitate a direct comparison of upper limits at $z = 1$, we adopt the single SMBH rate (dashed line) in the bottom-right panel of Figure~\ref{fig:rate_per_redshift}. The light blue line represents a linear uncertainty model, intersecting the pink curve at $\eta=1$ (the expected binary-driven TDE rate). In this model, we assume that only PSB galaxies host SMBH binaries, so the fraction of PSB galaxies at a given redshift directly sets the value of $\eta$. The purple bands at $z\lesssim1$ and $ z\sim2$ represent the PSB fractions from \cite{french_tidal_2016, french_host_2020}. 

Within this framework, the observed TDE rates in PSB galaxies align with binary-driven models for plausible values of $\eta$. For $z = 1$, the intersection of the PSB fraction (purple line) with the pink curve suggests that if PSBs host SMBH binaries, then the upper-limit TDE count at $\eta = 1$ would be reduced to $\sim 10^4$. While other processes may also contribute to TDE enhancement in PSBs, our results constrain the SMBH binary population required to explain current observations. 

Looking ahead, this framework is testable. Next-generation time-domain surveys, such as the Vera C. Rubin Observatory’s LSST and the Roman Space Telescope, are poised to transform our understanding of TDE demographics across mass and redshift \citep[e.g.,][]{spergel_wide-field_2015, ivezic_lsst_2019, gomez_identifying_2023}.
Rubin has already 
released its first images, revealing millions of galaxies and detailed views of the Milky Way and Virgo Cluster \citep{vera_c_rubin_ever-changing_2025}. With its unprecedented depth and cadence, Rubin is expected to detect thousands of TDEs, including those in previously under-sampled regions of parameter space. These surveys will enable statistically robust demographic studies, shedding light on host galaxy types, black hole masses, and the environments that favor TDE production. Our redshift-dependent predictions offer a falsifiable 
framework for interpreting future TDE observations and constraining the demographics of SMBH binaries.

\section{Acknowledgments}\label{sec:acknowledgements}
D.M. acknowledges the partial support from NSF graduate fellowship DGE-2034835, the UCLA Eugene V. Cota-Robles Fellowship, and the UCLA Mani L. Bhaumik Institute for Theoretical Physics Fellowship.
S.N. acknowledges the partial support of NSF-BSF grant AST-2206428, as well as Howard and Astrid Preston for their generous support. This research was supported in part by grant NSF PHY-2309135 to the Kavli Institute for Theoretical Physics (KITP). 

\appendix 
\section{Simulation Setup}\label{sec:appendix}


To model TDE formation in SMBH binaries, we use the Octupole-level Secular Perturbation Equations (OSPE) code \cite{naoz_hot_2011, naoz_resonant_2013, naoz_secular_2013, naoz_hot_2011}, which numerically solves the secular three-body equations of motion up to the octupole order term. This approach captures the long-term evolution of stellar orbits driven by the EKL mechanism while maintaining consistency with direct N-body simulations. OSPE operates under the secular approximation, averaging over orbital periods to focus on longer dynamical timescales. In this approximation, each orbit is treated as a massive wire with line density inversely proportional to the orbital velocity. These wires exert torques on one another, facilitating the exchange of angular momentum while conserving energy. 

We simulate 1,000 systems per SMBH mass configuration, incorporating the influence of two-body relaxation and the EKL mechanism. The mass of the disrupting SMBH is varied across $m_1 = [10^5, 10^6, 10^7]$~$ M_\odot$ while the binary mass ratio is fixed at $q = m_2/m_1 =  100$. The binary separation is fixed at half of the primary SMBH's sphere of influence, consistent with the expectation for bound binaries in post-merger galaxies \citep[][]{merritt_massive_2005}. The binary eccentricity is set at e$_{bin} = 0.5$, reflecting moderate eccentricities typical of post-merger SMBH pairs \citep[e.g.,][]{sesana_massive_2011}. 

We adopt a core-like stellar density profile with the power-law slope set to $\alpha = 1$, consistent with core-like profiles observed in PSBs \citep[][]{ stone_rates_2016}. Stellar masses are fixed at $m_\star = 0.8$ M$_\odot$ following \cite[][]{melchor_tidal_2024}. These choices reflect realistic conditions in post-merger environments, where SMBH binaries are most likely to form and dynamically influence TDE production. 
\section{Two-body Relaxation Rate}
The theoretical TDE rates from two-body relaxation , shown in Figure~\ref{fig:rate_per_mass}, are computed within the SMBH's sphere of influence, $R_{sph}$, as
\begin{equation}
    \Gamma_{single} (R_{sph}) \sim \frac{N_\star (R_{sph})}{t_{rlx}} \frac{1}{\ln (J_{circ}/J_{lc})} \ ,
\end{equation}\label{eq:gamma_rlx}

where $t_{rlx}$ is the relaxation time, $J_{circ}$ is the angular momentum of a circular orbit, and $J_{lc}$ is the angular momentum corresponding to the loss cone boundary. These are give by,
\begin{equation}
    J_{circ} = \sqrt{G M_{\rm BH} a} \ ,
\end{equation}

\begin{equation}
    J_{lc} = \sqrt{G M_{\rm BH} a (1-e^2)} \sim \sqrt{2 G M_{\rm BH} R_T} \ ,
\end{equation}
where $a$ is the semi-major axis and $R_T$ is the tidal disruption radius. Stars with angular momentum $J \leq J_{lc}$ are considered ``lost" to the SMBH, either through disruption or direct capture. Although these events occur close to the SMBH, the supply rate of stars is governed by stellar dynamics at larger radii within the nuclear star cluster. 
\subsection{Stellar Number Density}
 The stellar number density within the sphere of influence, $N_\star(R_{sph}$), is derived from the M$_{BH}-\sigma_\star$ relation using
\begin{equation}
    N_\star (r) = 2\frac{M_{\rm BH}}{m_\star} \left( \frac{G^2 \sqrt{M_{\rm BH} M_0}}{\sigma_0^2 r}\right)^{-3+\alpha} \ ,
\end{equation}
where $M_0=10^8~$M$_\odot$ and $\sigma_0=200~$km~sec$^{-1}$, are normalization constants. This profile assumes a spherical stellar distribution with density slope $\alpha$ and follows scaling relations observed in galactic nuclei (see Figure \ref{fig:Mh_vs_Mh}).

\subsection{TDE Parameters}
\begin{figure*}
   \begin{center} 
    \includegraphics[width=0.8\textwidth]{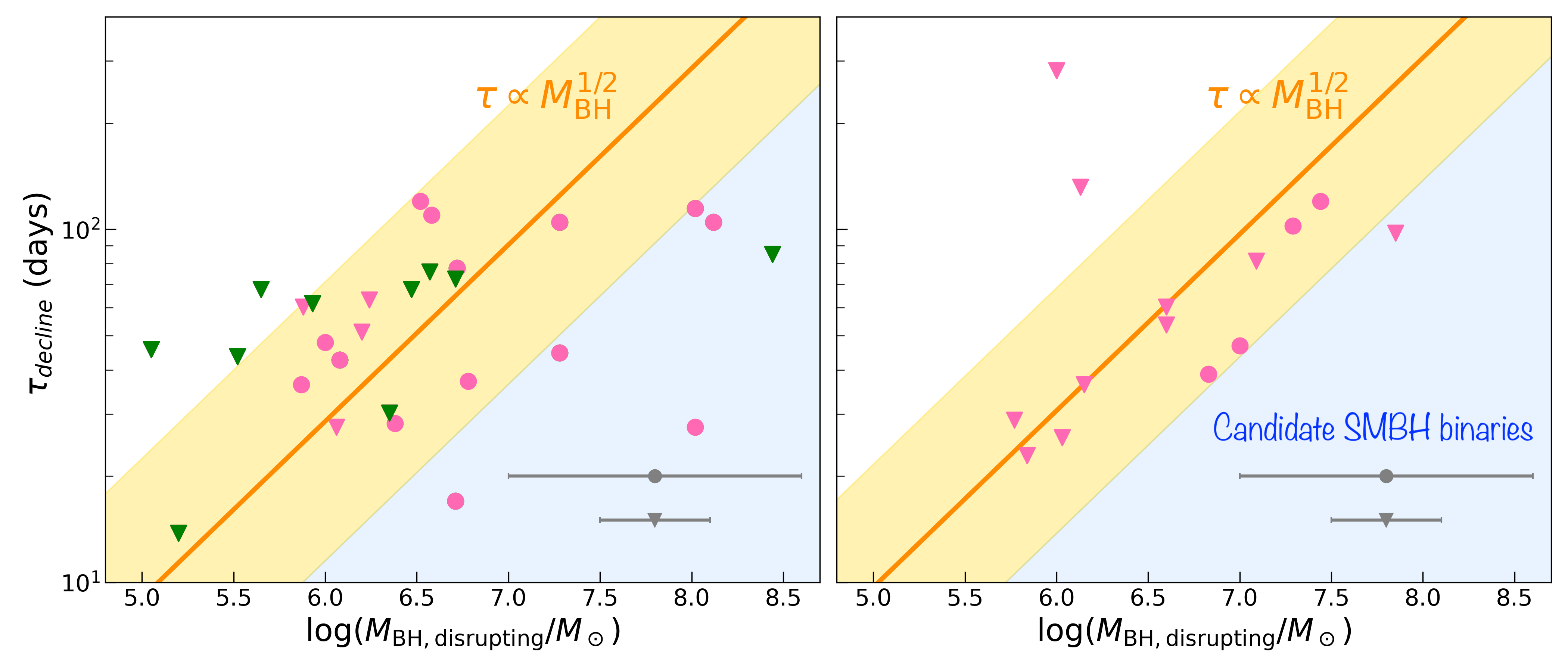} 
  \end{center}

  \caption{  \upshape {\bf Correlation between TDE light curves and black hole mass.} Scatter points show $\tau_{\rm decline}$, measured from the light curve decay timescale of individual TDEs, plotted against the estimated black hole mass ($M_{\rm BH}$) for each event. The blue shaded region indicates the parameter space where SMBH candidates may form TDEs. \textbf{Left panel:} TDE light curve properties adapted from \cite{hammerstein_final_2022}, TDE black hole masses adapted from \cite{mummery_fundamental_2024}. \textbf{Right panel:} TDE light curve properties adapted from \cite{van_velzen_late-time_2019}, TDE black hole masses adapted from \cite{mummery_fundamental_2024}. For both panels, the mass of the black hole is derived from $M_{\rm BH} - \sigma_\star$ when available, and from $M_{\rm BH} - M_{\rm gal}$; derivations can be found in \cite{mummery_fundamental_2024}. The orange line in both panels indicates the best-fit line with slope matching the $t_{\rm fb}$ theoretical line. The yellow range represents a $1\sigma$ deviation. In both plots, the systematic errors as stated in \cite{mummery_fundamental_2024} are shown in the corners: approximately $\sim 0.3$ dex and $\sim 0.8$ dex for $M_{\rm BH} - \sigma_\star$ (upside triangles) and $M_{\rm BH} - M_{\rm gal}$ (circles), respectively.}
  \label{fig:combined_tau_Mh}   
\end{figure*}

Table \ref{tab:yao_tde_data} lists 33 TDEs compiled from \cite{yao_tidal_2023}, including rise and decline half-times, SMBH mass estimates, and host galaxy velocity dispersions. Table \ref{tab:combined_tde_data} compiles decay timescales and corresponding SMBH/host galaxy parameters used in our comparison analyses (see Figure \ref{fig:combined_tau_Mh}).

\begin{table*}[t!]
\centering
\caption{Light curve and host galaxy properties of TDEs compiled by \cite{yao_tidal_2023}.}
\label{tab:yao_tde_data}
\begin{tabular}{lcccccr}
\toprule
  TDE Name &  $t_{1/2, \rm ~rise}$ &   $t_{1/2, \rm ~decline}$  &    $\log M_{{\rm BH}}$  & $\sigma_\star$ &  $\log M_{\rm gal}$ &  TDE Reference \\
  &  (days) & (days) &   (M$_{\odot})$ &   (km/s) &   (M$_{\odot}$) &  \\
\midrule
 AT2018iih &                31.0$^{+2.5}_{-1.5}$&                                   86.5$^{+3.3}_{-5.0}$ &                                   7.93 $\pm$                   0.35 &          148.64 $\pm $                14.42 &               $10.69^{+0.12}_{-0.16}$ &                          \cite{van_velzen_seventeen_2021}\\
 AT2018jbv &                34.4$^{+2.1}_{-1.4}$ &                                    65.9$^{+2.3}_{-1.7}$ &                                    6.77 $\pm$                    0.40 &               -                    &               $10.20^{+0.17}_{-0.19}$ &                         \cite{hammerstein_final_2023}           \\
 AT2018lna &                15.5$^{+1.3}_{-1.0}$ &                                   30.2$^{+1.3}_{-1.1}$ &                                   5.56 $\pm$                   0.51 &               -                     &               $ 9.50^{+0.12}_{-0.17}$ &                       \cite{van_velzen_seventeen_2021}\\
 AT2019baf &                23.2$^{+0.9}_{-1.0}$ &                                27.6$^{+0.6}_{-0.9}$ &                                   6.89 $\pm$                   0.24 &               -                     &              $ 10.27^{+0.04}_{-0.05}$ &                     
 \cite{yao_tidal_2023}    \\
 AT2019azh &                24.7$^{+1.3}_{-1.0}$ &                                    44.1$^{+1.1}_{-0.9}$ &                                   6.44 $\pm$                    0.33 &           67.99 $\pm $                2.03 &                $9.88^{+0.03}_{-0.03}$ &                       \cite{hinkle_discovery_2021}   \\
 AT2019bhf &                 9.9$^{+0.7}_{-0.9}$ &                                    29.1$^{+1.9}_{-1.4}$ &                                    7.10 $\pm $                   0.24 &               -                    &              $ 10.39^{+0.05}_{-0.06}$ &                      \cite{van_velzen_seventeen_2021}     \\
 AT2019cmw &                14.0$^{+0.3}_{-0.3}$ &                                    28.9$^{+0.7}_{-0.5}$ &                                    7.94 $\pm$                    0.42 &               -                    &              $ 10.88^{+0.17}_{-0.20}$ &                    
 \cite{yao_tidal_2023} \\
 AT2019dsg &                19.7$^{+2.3}_{-2.0}$ &                                   43.1$^{+1.0}_{-1.1}$ &                                    6.90 $\pm$                    0.32 &           86.89 $\pm $                3.92 &               $10.34^{+0.06}_{-0.05}$ &                       
 \cite{stein_tidal_2021}\\
 AT2019ehz &                15.7$^{+0.7}_{-0.8}$ &                                     28.0$^{+0.0}_{-1.0}$ &                                    5.81 $\pm $                   0.46 &               -                    &                $9.65^{+0.13}_{-0.16}$ &                       \cite{van_velzen_seventeen_2021}   \\
 AT2019qiz &                11.6$^{+0.3}_{-0.3}$ &                                     17.9$^{+0.7}_{-0.8}$ &                                    6.48 $\pm $                   0.33 &            69.7 $\pm $                 2.3 &               $10.28^{+0.04}_{-0.06}$ &                          
 \cite{nicholl_outflow_2020}\\
 AT2019vcb &                13.6$^{+1.1}_{-0.8}$ &                                   24.6$^{+0.4}_{-0.4}$ &                                   6.03 $\pm $                   0.36 &               -                    &               $ 9.77^{+0.03}_{-0.07}$ &                      
 \cite{hammerstein_final_2023}\\
  AT2020pj &                12.4$^{+0.7}_{-0.5}$ &                                    17.2$^{+1.3}_{-1.1}$ &                                     6.44 $\pm$                    0.31 &               -                    &             $  10.01^{+0.07}_{-0.08}$ &                      \cite{hammerstein_final_2023}   \\
 AT2020mot &                42.6$^{+1.3}_{-1.6}$&                                   46.1$^{+1.9}_{-2.1}$ &                                     6.66 $\pm $                   0.34 &           76.61 $\pm $               5.33 &             $  10.40^{+0.06}_{-0.08}$ &                    \cite{hammerstein_final_2023}   \\
 AT2020neh &                 6.4$^{+0.4}_{-0.4}$ &                                    16.4$^{+0.6}_{-0.6}$ &                                    5.43 $\pm$                    0.46 &            40.0 $\pm $                 6.0 &              $  9.80^{+0.05}_{-0.06}$ &                      \cite{angus_fast-rising_2022}\\
 AT2020ysg &                24.0$^{+2.1}_{-1.5}$ &                                    72.5$^{+2.1}_{-3.3}$ &                                     8.04 $\pm $                   0.33 &          157.78 $\pm $               13.03 &               $10.70^{+0.06}_{-0.07}$ &                    
 \cite{hammerstein_final_2023}\\
 AT2020vdq &                11.9$^{+1.7}_{-1.3}$ &                                  23.3$^{+1.5}_{-1.7}$ &                                   5.59 $\pm$                    0.37 &           43.56 $\pm $               3.07 &               $ 9.25^{+0.07}_{-0.11}$ &                     
 \cite{yao_tidal_2023}   \\
 AT2020vwl &                22.2$^{+0.8}_{-0.7}$ &                                27.4$^{+1.9}_{-1.7}$ &                                     5.79 $\pm$                    0.35 &           48.49 $\pm $                2.0 &               $ 9.89^{+0.08}_{-0.08} $&                     
 \cite{hammerstein_ztf_2021}\\
 AT2020wey &                13.9$^{+0.4}_{-0.4}$ &                                    5.2$^{+0.2}_{-0.2}$ &                                     5.40 $\pm $                   0.38 &           39.36 $\pm $                2.79 &               $ 9.67^{+0.09}_{-0.12}$ &                       
 \cite{arcavi_transient_2020}\\
 AT2020yue &                19.5$^{+1.0}_{-0.9}$ &                                   62.8$^{+2.0}_{-1.9}$ &                                  6.75 $\pm$                    0.32 &               -                    &               $10.19^{+0.10}_{-0.14}$ &                   
 \cite{yao_tidal_2023}   \\
AT2020abri &                16.7$^{+1.2}_{-0.9}$ &                                 31.7$^{+0.7}_{-0.8}$ &                                    5.62 $\pm $                   0.51 &               -                    &                $9.54^{+0.14}_{-0.17}$ &                    
\cite{yao_tidal_2023}  \\
AT2020acka &                26.9$^{+1.6}_{-1.8}$ &                                    28.8$^{+0.7}_{-0.5}$ &                                    8.23 $\pm$                    0.40 &          174.47 $\pm $                 25.3 &              $ 11.03^{+0.15}_{-0.19}$ &                         
\cite{hammerstein_ztf_2021}\\
 AT2021axu &                23.9$^{+0.5}_{-0.6}$ &                                   33.4$^{+0.9}_{-1.0}$ &                                    6.59 $\pm$                    0.55 &            73.5 $\pm $                17.26 &               $10.20^{+0.11}_{-0.13}$ &                 
 \cite{hammerstein_ztf_2021}\\
 AT2021crk &                10.2$^{+0.7}_{-0.4}$ &                                  20.9$^{+1.1}_{-1.1}$ &                                   6.12$ \pm$                    0.39 &           57.62 $\pm $                 6.29 &               $ 9.89^{+0.11}_{-0.10}$ &                    
 \cite{yao_tidal_2023}               \\
 AT2021ehb &                23.7$^{+1.9}_{-1.4}$ &                                  50.5$^{+3.6}_{-3.8}$ &                                    7.16 $\pm $                   0.32 &           99.58 $\pm $                3.83 &               $10.23^{+0.01}_{-0.02}$ &                       \cite{yao_tidal_2022}
 \\
 AT2021jjm &                 9.1$^{+0.7}_{-0.7}$ &                                    29.1$^{+2.6}_{-1.7}$ &                                    5.51 $\pm$                    0.51 &               -                    &               $ 9.47^{+0.13}_{-0.14}$ &                        
 \cite{yao_ztf_2021}\\
 AT2021mhg &                17.2$^{+0.7}_{-0.7}$ &                                    14.7$^{+1.1}_{-1.0}$ &                                     6.13 $\pm $                   0.37 &           57.78 $\pm $                5.25 &                $9.65^{+0.12}_{-0.14}$ &                          
 \cite{chu_ztf_2021}\\
 AT2021nwa &                27.1$^{+0.6}_{-0.8}$ &                                    76.2$^{+1.9}_{-1.6}$ &                                     7.22 $\pm$                    0.32 &          102.44 $\pm $                5.37 &             $  10.13^{+0.03}_{-0.05}$ &                       
 \cite{yao_ztf_2021}\\
 AT2021qth &                15.8$^{+1.2}_{-1.3}$ &                                    39.1$^{+1.3}_{-2.0}$ &                                     5.95 $\pm $                   0.48 &               -                     &              $  9.73^{+0.14}_{-0.21}$ &                    
 \cite{yao_tidal_2023}              \\
 AT2021sdu &                12.2$^{+0.4}_{-0.4}$ &                                  11.0$^{+0.3}_{-0.4}$ &                                  6.68 $\pm $                  0.29 &               -                   &             $  10.15^{+0.07}_{-0.09}$ &                     
 \cite{chu_ztf_2021}\\
 AT2021uqv &                14.9$^{+0.7}_{-0.7}$ &                                  36.0$^{+2.2}_{-2.0}$ &                                    6.27$ \pm$                   0.39 &            62.3 $\pm $                7.08 &               $10.14^{+0.08}_{-0.11} $&                            \cite{yao_ztf_2021}
 \\
 AT2021utq &                14.6$^{+0.6}_{-0.6}$ &                                  43.4$^{+5.8}_{-4.3}$ &                                    5.84 $\pm$                   0.43 &               -                   &               $ 9.66^{+0.09}_{-0.12}$ &                     
 \cite{yao_tidal_2023}                \\
 AT2021yzv &                51.8$^{+1.4}_{-1.2}$ &                                    69.9$^{+2.6}_{-2.6}$ &                                     7.90 $\pm$                    0.40 &          146.38 $\pm $                20.78 &               $10.83^{+0.12}_{-0.15}$ &                            
 \cite{chu_ztf_2022}\\
 AT2021yte &                18.4$^{+0.5}_{-0.6}$ &                                   23.7$^{+0.7}_{-0.7}$ &                                     5.13 $\pm$                   0.45 &           34.22 $\pm $               4.81 &               $ 9.17^{+0.17}_{-0.21}$ &                              \cite{yao_ztf_2021}       \\
\bottomrule

\end{tabular}
\end{table*}



\begin{table*}[t!]
\centering
\caption{Decay timescales for TDEs from \cite{hammerstein_integral_2023} and \cite{van_velzen_first_2019} and SMBH/host properties for TDEs from \cite{mummery_fundamental_2024}.}
\label{tab:combined_tde_data}
\begin{tabular}{lcccccr}
\toprule
TDE Name & $\tau$ & $\sigma_\star$  &  M$_{\rm BH}$ from $\sigma_\star$   &  M$_{\rm gal}$ &  M$_{\rm BH}$ from M$_{\rm gal}$ & TDE Reference \\
 &log(day) & (km/s)  & (M$_\odot$)  &(M$_\odot$) & (M$_\odot$) &  \\
\toprule
  AT2018fyk &$1.99^{0.02}_{0.02}$&            158.0 &                7.85 &      10.61 &                  7.65 & \cite{wevers_evidence_2019}\\
  AT2018dyb &$1.59^{0.01}_{0.01}$&              -&                 -&      10.10 &                  6.83 & \cite{leloudas_spectral_2019}\\
  AT2018zr &        $1.83^{0.03}_{0.03}$ &             50.0 &                5.65 &      10.03 &                  6.71 & \cite{holoien_ps18kh_2019} \\
   & &               &                 &         &                    &    \cite{van_velzen_first_2019} \\
  AT2018hco &        $2.04^{0.02}_{0.02}$ &              -&                 -&       9.95 &                  6.58 &\cite{van_velzen_optical-ultraviolet_2020}\\
  AT2018iih &        $2.06^{0.03}_{0.03}$ &              -&                 -&      10.84 &                  8.02 & \cite{van_velzen_optical-ultraviolet_2020} \\
  AT2018hyz &        $1.71^{0.01}_{0.01}$ &             67.0 &                6.20 &       9.53 &                  5.91 &  \cite{short_tidal_2020}\\
  & &               &                 &         &                    &    \cite{van_velzen_optical-ultraviolet_2020} \\
  AT2018lni & $1.78^{0.02}_{0.02}$&             56.0 &                5.88 &       9.84 &                  6.41 & \cite{van_velzen_optical-ultraviolet_2020} \\
  AT2018lna &        $1.66^{0.02}_{0.02}$ &             36.0 &                5.05 &       9.41 &                  5.70  & \cite{van_velzen_optical-ultraviolet_2020}\\
  AT2018jbv &$2.02^{0.03}_{0.03}$&              -&                 -&      10.38 &                  7.28 & \cite{hammerstein_final_2023} \\
  AT2019ahk &$1.67^{0.01}_{0.01}$&              -&                 -&      10.21 &                  7.00  & \cite{holoien_discovery_2019}\\
  AT2019cho &$1.89^{0.04}_{0.03}$&              -&                 -&      10.04 &                  6.72 &  \cite{van_velzen_optical-ultraviolet_2020}\\
  AT2019bhf &$1.65^{0.03}_{0.03}$&              -&                 -&      10.38 &                  7.28  & \cite{van_velzen_optical-ultraviolet_2020}\\
  AT2019azh &$1.8^{0.02}_{0.02}$&             68.0 &                6.24 &       9.79 &                  6.32 &  \cite{van_velzen_optical-ultraviolet_2020} \\
    & &               &                 &         &                    &    \cite{hinkle_discovery_2021} \\
        & &               &                 &         &                    &    \cite{liu_uvoptical_2022} \\
  AT2019dsg &$1.86^{0.02}_{0.02}$&             87.0 &                6.71 &      10.54 &                  7.54 & \cite{van_velzen_optical-ultraviolet_2020} \\
  AT2019ehz &$1.64^{0.01}_{0.01}$&             47.0 &                5.52 &       9.69 &                  6.17 &\cite{van_velzen_optical-ultraviolet_2020}\\
  AT2019mha &$1.23^{0.03}_{0.03} $ &              -&                 -&      10.03 &                  6.71  & \cite{van_velzen_optical-ultraviolet_2020}\\
  AT2019meg &$1.68^{0.02}_{0.02}$&              -&                 -&       9.59 &                  6.00  & \cite{van_velzen_optical-ultraviolet_2020}\\
  AT2019lwu &$1.45^{0.03}_{0.03}$&              -&                 -&       9.83 &                  6.38  & \cite{van_velzen_optical-ultraviolet_2020} \\
  AT2019qiz &$1.48^{0.01}_{0.01}$&             72.0 &                6.35 &      10.02 &                  6.70 & \cite{van_velzen_optical-ultraviolet_2020}\\
          & &               &                 &         &                    &    \cite{nicholl_outflow_2020} \\
  AT2019teq &$2.08^{0.06}_{0.06}$&              -&                 -&       9.91 &                  6.52 & \cite{hammerstein_final_2023} \\
  AT2019vcb &$1.56^{0.01}_{0.01}$&              -&                 -&       9.51 &                  5.87 &  \cite{hammerstein_final_2023}\\
  AT2020pj &$1.57^{0.02}_{0.02}$&              -&                 -&      10.07 &                  6.78 & \cite{hammerstein_final_2023} \\
  AT2020ddv &$1.79^{0.02}_{0.02}$&             58.0 &                5.93 &      10.22 &                  7.01  &\cite{hammerstein_final_2023}\\
  AT2020ocn &$1.88^{0.03}_{0.03}$&             81.0 &                6.57 &      10.32 &                  7.17  & \cite{hammerstein_final_2023}\\
  AT2020mot &$1.83^{0.01}_{0.01}$&             77.0 &                6.47 &      10.28 &                  7.10 & \cite{hammerstein_final_2023} \\
  AT2020mbq &$1.63^{0.02}_{0.01}$&              -&                 -&       9.64 &                  6.08  & \cite{hammerstein_final_2023}\\
  AT2020qhs &$1.93^{0.02}_{0.02}$&            215.0 &                8.44 &      11.33 &                  8.81  & \cite{hammerstein_final_2023}\\
  AT2020riz &$1.44^{0.02}_{0.02}$&              -&                 -&      10.84 &                  8.02  & \cite{hammerstein_final_2023}\\
  AT2020wey &$1.14^{0.02}_{0.02}$&             39.0 &                5.20 &       9.85 &                  6.41 & \cite{hammerstein_final_2023} \\
  AT2020zso &$1.44^{0.03}_{0.03}$&             62.0 &                6.06 &      10.16 &                  6.91  & \cite{wevers_elliptical_2022}\\
          & &               &                 &         &                    &    \cite{ hammerstein_final_2023} \\
  AT2020ysg &$2.02^{0.02}_{0.02}$&              -&                 -&      10.90 &                  8.12  & \cite{hammerstein_final_2023}\\
  SDSS-TDE2 &$2.08^{0.09}_{0.08}$&              -&                 -&      10.48 &                  7.44  & \cite{van_velzen_optical_2011}\\
   PTF-09ge &$1.78^{0.02}_{0.02}$&             82.0 &                6.60 &      10.33 &                  7.20  & \cite{arcavi_continuum_2014}\\
  PTF-09axc &$2.45^{0.39}_{0.43} $    &         60.0 &                6.00 &      10.07 &                  6.78  & \cite{arcavi_continuum_2014}\\
  PTF-09djl &$2.12^{0.52}_{0.33}$&             64.0 &                6.13 &       9.93 &                  6.56  & \cite{arcavi_continuum_2014}\\
   PS1-10jh &$1.56^{0.03}_{0.03}$&             65.0 &                6.15 &       9.60 &                  6.02  & \cite{gezari_ultravioletoptical_2012}\\
ASASSN-14ae &$1.46^{0.01}_{0.01}$&             53.0 &                5.77 &       9.90 &                  6.50  & \cite{holoien_asassn-14ae_2014}\\
  iPTF-15af &$1.91^{0.03}_{0.02}$&            106.0 &                7.09 &      10.35 &                  7.23  & \cite{blagorodnova_broad_2019}\\
ASASSN-15oi &$1.41^{0.01}_{0.01}$&             61.0 &                6.03 &      10.07 &                  6.77  & \cite{holoien_asassn-14ae_2014}\\
  OGLE16aaa &$2.01^{0.06}_{0.05}$&              -&                 -&      10.39 &                  7.29  & \cite{wyrzykowski_ogle16aaa_2016}\\
 iPTF-16axa &$1.73^{0.01}_{0.01}$&             82.0 &                6.60 &      10.24 &                  7.05  & \cite{hung_revisiting_2017}\\
 iPTF-16fnl &$1.36^{0.01}_{0.01}$     &        55.0 &                5.84 &       9.50 &                  5.86 & \cite{blagorodnova_iptf16fnl_2017} \\

\bottomrule
\end{tabular}
\end{table*}

\bibliography{TDE_demographics_references}{}
\bibliographystyle{aasjournal}



\end{document}